%% file: curved-ring.tex
\documentclass[aps,prb,reprint,showpacs,superscriptaddress]{revtex4-2}
\usepackage{amsmath}
\usepackage{amssymb}
\usepackage{amsthm}
\usepackage{bbm}
\usepackage{graphicx}
\usepackage{hyperref}
\usepackage{physics}
\usepackage{systeme}
\usepackage{times}
\usepackage{epstopdf}
\usepackage{float}

\hypersetup{
  colorlinks=true,linkcolor=blue,citecolor=blue,
  filecolor=blue,urlcolor=blue,breaklinks=true
}

\begin{document}

\title{Study of electronic properties, Magnetization and persistent currents in a mesoscopic ring by controlled curvature}

\author{Lu\'{i}s Fernando C. Pereira}
\email{luisfernandofisica@hotmail.com}
\affiliation{
        Departamento de F\'{i}sica,
        Universidade Federal do Maranh\~{a}o,
        65085-580, S\~{a}o Lu\'{i}s, Maranh\~{a}o, Brazil
      }

\author{Fabiano M. Andrade}
\email{fmandrade@uepg.br}
\affiliation{
        Departamento de Matem\'{a}tica e Estat\'{i}stica,
        Universidade Estadual de Ponta Grossa,
        84030-900 Ponta Grossa, Paran\'{a}, Brazil
      }

\author{Cleverson Filgueiras}
\email{cleverson.filgueiras@dfi.ufla.br}
\affiliation{
        Departamento de F\'{i}sica,
        Universidade Federal de Lavras, Caixa Postal 3037,
        37200-000, Lavras, Minas Gerais, Brazil
      }

\author{Edilberto O. Silva}
\email{edilbertoo@gmail.com}
\affiliation{
        Departamento de F\'{i}sica,
        Universidade Federal do Maranh\~{a}o,
        65085-580, S\~{a}o Lu\'{i}s, Maranh\~{a}o, Brazil
      }

\date{\today }

\begin{abstract}
We study the model of a noninteracting spinless electron gas confined to the two-dimensional localized surface of a cone in the presence of external magnetic fields. The localized region is characterized by an annular radial potential. We write the Schr\"{o}dinger equation and use the thin-layer quantization procedure to calculate the wavefunctions and the energy spectrum. In such a procedure, it arises a geometry induced potential, which depends on both the mean and the Gaussian curvatures. Nevertheless, since we consider a ring with a mesoscopic size, the effects of the Gaussian curvature on the energy spectrum are negligible. The magnetization and the persistent current are analyzed. In the former, we observed the Aharonov-Bohm (AB) and de Haas-van Alphen (dHvA) types oscillations. In the latter, it is observed only the AB type oscillations. In both cases, the curvature increases the amplitude of the oscillations.
\end{abstract}

\pacs{73.63.Kv,73.23.-b,73.63.-b,74.78.Na}
\maketitle

\section{Introduction}

\label{intro}

Quantum Rings are systems of great interest in condensed matter physics. This is due to the numerous physical phenomena that are observed in these structures. For example, we can analyze the dependence of resistance, which constitutes a case of transport property, with the external magnetic field \cite{PRL.1985.55.1610,PRB.1993.48.15148,PRB.1996.53.6947,PRB.2007.76.03510}, where the oscillations observed in the experiments correspond to the analog of the AB effect \cite{PR.1959.115.485}. Studies concerning optical properties are also analyzed in these optical mesoscopic structures. For example, in the Ref. \cite{SM.2013.58.94}, authors study photoionization cross section in a two-dimensional quantum ring. The optical properties of an  exciton in a two-dimensional ring threaded by a magnetic flux are study in the Ref. \cite{PRB.2011.84.235103}. Another field of physics widely studied in mesoscopic rings corresponds to the thermodynamic properties such as
the persistent current and magnetization.

Both the persistent current and magnetization are results of the application of external magnetic fields. The simple model $1$D ring threaded by a magnetic flux explains the origin of the persistent currents \cite{PRL.7.46.1961,PLA.1983.96.365}. In this case, the boundary condition is modified by the gauge, which cause the periodicity of the energy levels and other physical quantities. For example, the persistent current which, according to the Byers-Yang relation \cite{PRL.7.46.1961}, it is given by the magnetic flux derivative of free energy of the ring and, so is an equilibrium property of the ring.
For the $1$D ring model, the magnetization is proportional to the persistent current. As observed by Tan and Inkson \cite{PRB.1999.60.5626}, this proportionality between persistent current and magnetization remains even for a $2$D ring model, as long as the study interval is weak magnetic fields. The AB and dHvA oscillations are observed when the magnetic field varies.
In the ideal case of zero temperature, the amplitude of these oscillations is maximal. Nevertheless, non-zero temperatures decrease the signal of persistent current and magnetization. This result from the sensitivity of the mesoscopic systems to the increasing of the temperature \cite{PRL.1993.70.2020,PRB.1988.37.6050,PRL.1989.62.587}.

Modern experimental techniques allow to make of non-planar objects of reduced dimensions \cite{PE.2000.6.828,Nano.2001.12.399,JACS.2005.127.13782}.This motivated the study of physical phenomena in mesoscopic systems that present curvature \cite{APL.2006.88.212113,PRB.2007.75.205309}. The method employed to address such a system
follows the well-known thin-layer quantization procedure employed by da Costa \cite{PRA.1981.23.1982}, and widely used in the literature \cite{PRB.2004.69.195313,EPL.2007.79.57001,PRA.2018.98.062112,PRA.2017.96.022116,PRA.2014.90.042117,AoP.2016.364.68,PRL.2008.100.230403,PRB.2013.87.174413,PRL.2014.112.257203,PRL.2010.105.206601,JMP.2019.60.023502,EJP.2016.38.015405,NTch.2016.27.135302,PE.2019.106.200,PRB.2018.97.241103}. An interesting result that emerges from the thin-layer quantization procedure is that a particle on a curved surface experiences a potential of purely geometric origin which depends on the mean and Gaussian curvatures. Thus, even in the absence of interactions of any nature, the particles cannot move
around freely on the surface.

The study of defects in solids is also an important research area in physics. A widely used approach is based on the Riemann-Cartan, which involves nontrivial metric and torsion \cite{KATANAEV.2005}. A linear defect, for instance disclination or dislocation, changes the topology of a medium \cite {PLA.1994.195.90}. By using Volterra process \cite{KLEMAN.1977}, the disclination is obtained by either removing or inserting a wedge of material. This procedure makes the space present positive or negative curvature. In the case of dislocations, the defect is associated with the appearance of torsion.

In this paper, we are interested in studying the physical implications caused by curvature effects on the motion of a noninteracting spinless electron gas bound to a two-dimensional localized region in the presence of external magnetic fields. We use the model proposed by Tan-Inkson to constrain the motion of the electrons to this two-dimensional localized region \cite{SST.1996.11.1635}. By employing the thin-layer quantization procedure, we obtain the Schr\"odinger equation \cite{PRL.2008.100.230403}, and solved it to find the energies and wave functions of the electrons on the two-dimensional localized curved surface. We rigorously analyze the impacts of curvature on the energies. We also study the zero temperature Fermi energy, determine the expressions for magnetization and persistent current and investigate the main physical consequences caused by curvature effects and compare them with results in the literature.

\section{Description of the model}
\label{sec:model}

We are interested here in studying the motion of a spinless charged
particle constrained to move on a curved surface in the presence of
magnetic fields.
We employ the procedure of \cite{PRL.2008.100.230403} for studying
the quantum mechanics of a constrained particle, which is based on
da Costa's thin-layer quantization procedure  \cite{PRA.1981.23.1982}.
In  \cite{PRL.2008.100.230403},
by making a proper choice of the gauge, it was shown that the dynamics is exactly
separable in two components: a surface and a transverse dynamics.
In the transverse motion, the dynamics is described by a
one-dimensional Schr\"{o}dinger equation with a transverse potential,
whereas the motion on the surface is described by a two-dimensional
Schr\"odinger equation in which appears a geometric potential,
given in terms of the mean and the Gaussian curvatures.
The geometric potential stems from the two-dimensional
confinement on the surface. In our study, we are interested only in the motion on the surface.

Let us consider, therefore, a non-interacting 2DEG constrained to
move on a curved surface in the presence of both a magnetic field
and a radial potential \cite{SST.1996.11.1635} given by
\begin{equation}
	V\left( r\right) =\frac{a_{1}}{r^{2}}+a_{2}r^{2}-V_{0},  \label{pot.radial}
\end{equation}
with $V_{0}=2\sqrt{a_{1}a_{2}}$.
This radial potential has a minimum at
$r_{0}=\left( a_{1}/a_{2}\right) ^{1/4}$.
For $r\rightarrow r_{0}$, we obtain the parabolic potential model,
\begin{equation}
V_{par}(r) \approx \frac{\mu \omega_{0}^{2}}{2}(r-r_{0})^{2}, \label{pot.parab}
\end{equation}
where
$\omega_{0}=\sqrt{8a_{2}/\mu }$ characterizes the strength
of the transverse confinement.
The potential (\ref{pot.radial}) describes a 2D quantum ring,
nevertheless, it can describe others physical systems.
For example, if $a_{1}=0$, we have a quantum dot and if $a_{2}=0$, we
have a quantum anti-dot.
Both the radius and the width of the ring can be adjusted independently
by suitable choices of $a_{1}$ and $a_{2}$.

In this work, the curved surface is defined by the following line element
in polar coordinates  \cite{AoP.2008.323.3150,JMP.2012.53.122106}
\begin{equation}
	\label{eq:line_element}
	ds^{2}=dr^{2}+\alpha ^{2}r^{2}d\theta ^{2},
\end{equation}
with $r\geq 0$ and $0\leq \theta <2\pi $. For $0<\alpha <1$ (deficit
angle), the metric above describes an actual conical surface,
while for $\alpha >1$ (proficit angle), it represents a saddle-like
surface.
In what follows we focus our analysis in a conical surface, in which
$0 < \alpha \leq 1$ ($\alpha=1$ leads to a flat surface). In this case, the Gaussian and the mean curvatures are
given, respectively, by \cite{EPL.2007.80.46002}
\begin{equation}
	\mathcal{K}=
	\left( \frac{1-\alpha }{\alpha }\right) \frac{\delta (r)}{r},
	\qquad
	\mathcal{H}=\frac{\sqrt{1-\alpha ^{2}}}{2\alpha r}.  \label{curvat}
\end{equation}
The corresponding geometric potential is written as
\begin{equation}
	V_{g}(r)=-\frac{\hbar ^{2}}{2\mu}\left[ \frac{(1-\alpha ^{2})}{4\alpha
		^{2}r^{2}}-\left( \frac{1-\alpha }{\alpha }\right) \frac{\delta (r)}{r}
	\right],  \label{Vgeo}
\end{equation}
with $\mu$ is the electron effective mass.
For the field configuration, we consider a superposition of two magnetic
fields, $\mathbf{B}=\mathbf{B}_{1}+\mathbf{B}_{2}$, with
$\mathbf{B_{1}}=B\mathbf{\hat{z}}$ being a uniform magnetic field and
$\mathbf{B_{2}}=(l\hbar/e\alpha r)\delta(r)\mathbf{\hat{z}}$ being a
magnetic flux tube, with $l=\varPhi/\varPhi_{0}$ being the AB
flux parameter, $e$ is the electric charge, and $\varPhi_{0}=h/e$ is the
magnetic flux quantum.
The field $\mathbf{B}$ is obtained from the vector potential $\mathbf{A}=\mathbf{A}_{1}+\mathbf{A}_{2}$, where $\mathbf{A}_{1}=(Br/2\alpha)\boldsymbol{\hat{\varphi}}$ and $\mathbf{A}_{2}=(l\hbar/e\alpha r)\boldsymbol{\hat{\varphi}}$.

Since we are only interested in the dynamics on the surface, we ignore
the transverse one. Thus, the relevant equation is
\begin{equation}
	H\chi _{S}\left( r,\varphi \right) =E\chi_{S}\left( r,\varphi \right), \label{chr}
\end{equation}
where
\begin{align}
H = {} & -\frac{\hbar ^{2}}{2\mu }\left[ \frac{1}{r}\frac{\partial }{\partial r}
\left( r\frac{\partial }{\partial r}\right) +\frac{1}{\alpha ^{2}r^{2}}
\left( \frac{\partial }{\partial \varphi }-il\right) ^{2}\right]\notag  \\
&-\frac{\hbar ^{2}}{2\mu }\left[ \frac{ieB}{\hbar \alpha ^{2}}\left( \frac{
	\partial }{\partial \varphi }-il\right) +\frac{e^{2}B^{2}r^{2}}{4\hbar
	^{2}\alpha ^{2}}\right]\notag  \\
&-\frac{\hbar ^{2}}{2\mu}\left[ \frac{(1-\alpha ^{2})}{4\alpha
	^{2}r^{2}}-\left( \frac{1-\alpha }{\alpha }\right) \frac{\delta (r)}{r}
\right]\notag
\\
& +\frac{a_{1}}{r^{2}}+a_{2}r^{2}-V_{0}.  \label{hamiltonian}
\end{align}
By considering solutions of the form
\begin{equation}
\chi _{S}(r,\varphi )=e^{im\varphi }f_{m}\left( r\right)
\qquad m = 0, \pm 1, \pm 2, \ldots  \label{ansatz}
\end{equation}
equation (\ref{hamiltonian}) results in the following radial equation for a specific quantum number $m$:
\begin{equation}
-\frac{\hbar ^{2}}{2\mu }\frac{1}{r}\frac{d}{dr}\left( r\frac{d}{dr}\right)
f_{m}(r)+V_{\rm eff}f_{m}(r)=E_{m}f_{m}(r),\label{radial_equation}
\end{equation}
where
\begin{equation}
V_{\rm eff}=\frac{\hbar ^{2}}{2\mu }\left( \frac{L^{2}}{r^{2}}+\frac{r^{2}}{
	4\lambda ^{4}}-\kappa ^{2}\right) + \frac{\hbar ^{2}}{2\mu}\left( \frac{1-\alpha }{\alpha }\right) \frac{\delta (r)}{r} \label{V-efetivo}
\end{equation}
is the effective induced potential.  We define the effective angular momentum, the effective cyclotron frequency, the effective magnetic length and a constant parameter, respectively, as
\begin{equation}
L=\sqrt{\left( \frac{m-l}{\alpha }\right) ^{2}+\frac{2\mu a_{1}}{\hbar ^{2}}-
	\frac{1-\alpha ^{2}}{4\alpha ^{2}}}\;,  \label{Mom.Angular}
\end{equation}
\begin{equation}
\omega =\sqrt{\left( \frac{\omega _{c}}{\alpha }\right) ^{2}+\omega _{0}^{2}}\;,  \label{Freq.ciclot.efet}
\end{equation}
\begin{equation}
\lambda =\sqrt{\frac{\hbar }{\omega \mu }}\;,  \label{comp.magnet}
\end{equation}
\begin{equation}
\kappa ^{2}=\frac{\left( m-l\right) \mu \omega _{c}}{\hbar \alpha ^{2}}+
\frac{2\mu V_{0}}{\hbar ^{2}}.  \label{Kappa}
\end{equation}
In Eq. (\ref{Freq.ciclot.efet}), $\omega_{c}=eB/\mu$ is the cyclotron
frequency and $\omega_{0}$ was defined above. It is
important to emphasize that the presence of the $\delta$ function in
Eq. (\ref{radial_equation}), which is a short-range potential, suggests
that singular solutions should be taken into account in this
approach. According to von Newman's theory of self-adjoint extensions \cite{Book.1975.Reed.II}, the necessity of inclusion of the irregular solutions of Eq. (\ref{radial_equation})
steams from the fact that $H$ (Eq. \ref{hamiltonian}) is not self-adjoint for $\left\vert L\right\vert <1$
\cite{PRD.2012.85.041701,AoP.2013.339.510}. However, for mesoscopic
effective angular moment (\ref{Mom.Angular}) is much larger than $1$, so
that there is no value of $L$ belonging to the range required above. Due
to that, we can ignore the $\delta$ function in the radial equation
(\ref{V-efetivo}) and we consider only the regular solution at $r=0$.

In this manner, the energy eigenvalues and wavefunctions of
Eq. (\ref{chr}) are given by
\begin{equation}
E_{n,m}=\left( n+\frac{1}{2}+\frac{L}{2}\right) \hbar \omega -\frac{\left(
	m-l\right) \hbar \omega _{c}}{2\alpha ^{2}}-V_{0}
	\label{Enm_ring}
\end{equation}
and
\begin{align}
\chi_{n,m}(r,\varphi )= {}
&\frac{1}{\lambda}\sqrt{\frac{\Gamma (L+n+1)}{2^L n!\,\Gamma (L+1)^2}}e^{ -\frac{r^2}{4\lambda^2}}e^{im\varphi }\left(\frac{r}{\lambda}\right)^L \notag \\
&\times \, {_{1}F_{1}}\left(-n, 1+L, \frac{r^2}{2\lambda^2}\right). \label{solution}
\end{align}
In addition, the effective radius $r_{n,m}$ of the states with quantum number $m$ is given by
\begin{equation}
r_{n,m}=\left( 2L\right) ^{\frac{1}{2}}\lambda  \label{raio_nm},
\end{equation}
where $L$ and $\lambda$ are given by (\ref{Mom.Angular}) and (\ref{comp.magnet}), respectively.
\section{Electronic States}

In this section, we analyze the energy spectrum (\ref{Enm_ring}). In the numerical analysis, we consider a $2$D ring defined by Eq. (\ref{pot.radial}), with $a_{1}=1.8154 \times 10^{9}$ meV $nm^{2}$  and
$a_{2}=5.4654\times 10^{-4}$ meV $nm^{-2}$, which describe a mesoscopic ring of average radius $r_{0}=1350$ nm. The sample is made of GaAs, so that the electron effective mass is $\mu=0.067\mu_{e}$, where $\mu_{e}$ is the electron mass.

\begin{figure}[!ht!]
	\centering
	\includegraphics[width=\columnwidth]{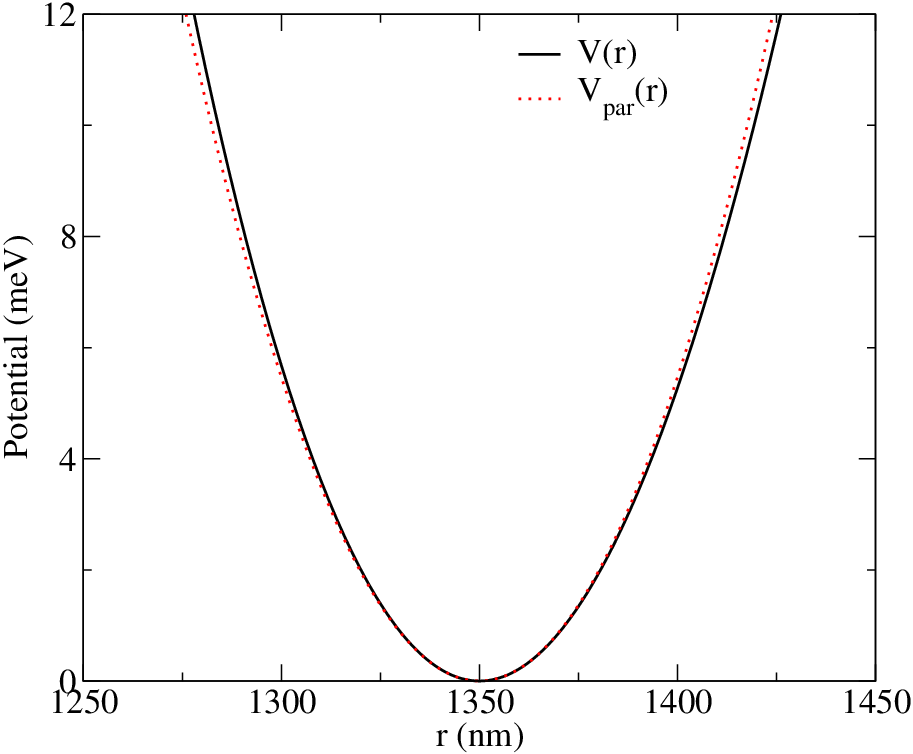}
	\caption{Sketch of the radial potential (solid black line) and the
	parabolic potential (solid red line). The radial potential describes
	a ring of average radius $r0=1350$ nm, while the parabolic potential is defined by $\hbar \omega _{0}=2.23$ meV.}
	\label{Potencial}
\end{figure}

In the Fig. (\ref{Potencial}), we have plotted the radial potential defined by Eq. (\ref{pot.radial}). In addition, the parabolic potential (Eq. (\ref{pot.parab})) is plotted, where the strength of the transverse confinement is given by $\hbar \omega _{0}=2.23$ meV. As we can see, the parabolic and radial models do not have a significant difference in the energy interval considered.

Using the Eq. (\ref{Enm_ring}), we can show that all
subband minimum lie at the same $m_{0}$ value given by
\begin{equation}
m_{0}=\frac{eBr_{0}^{2}}{2\hbar }\sqrt{1-\frac{1}{a_{1}}\frac{\hbar ^{2}}{2\mu }\frac{1-\alpha ^{2}}{4\alpha ^{2}}.}  \label{minimo.curvo}
\end{equation}
Therefore, when the magnetic field is null, the subband minima is $m=0$. The variation of the magnetic field changes the minimal subbands. For $\alpha=1$, we recover the result obtained by Tan e Inkson, which correspond to the number of quantum flux threading a ring with an effective radius $r_{0}$.

\begin{figure}[!ht!]
	\centering
	\includegraphics[width=\columnwidth]{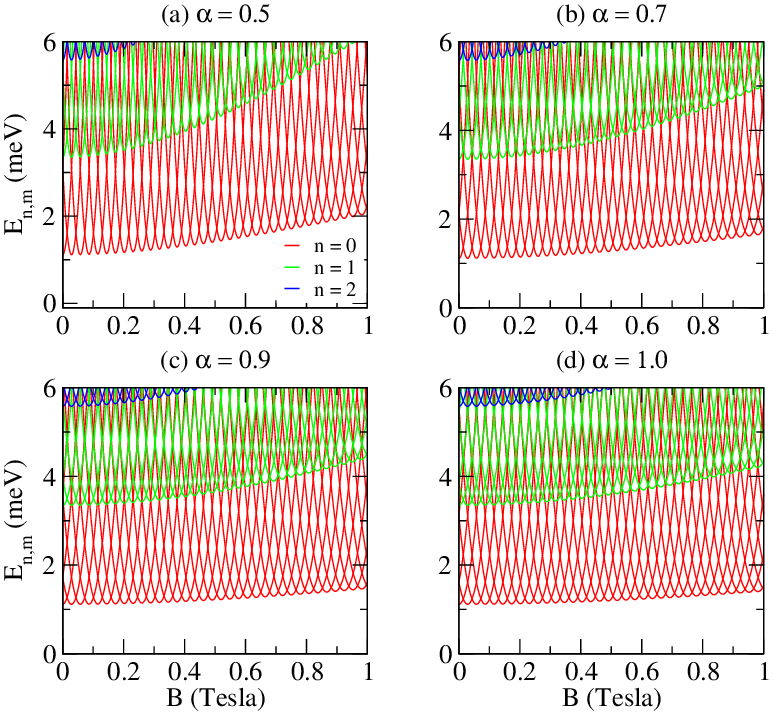}
	\caption{The energy levels of a quantum ring as a function of the magnetic field for different curvature values.}
	\label{Energia_B}
\end{figure}

If the magnetic field is null, the subbands are symmetric with respect to $m=0$. The states with $m \neq 0$ are doubly degenerate. The double degeneration is due to the rotational symmetry present in the ring.
The symmetry with respect to states in subband different is broken due to the potential of the antidot. The energy separation between neighboring subbands is given by  $\hbar \omega _{0}$, and does not depend on the value of the parameter $\alpha$. If the magnetic field is different from zero, the energy separation between neighboring is $\hbar \omega$. Independently of the presence or not of an applied, the curvature has the role of increasing the energy of the states. This makes the number states decrease in the energy interval. For the physical phenomena that we will deal in the next sections, the number of electrons in the system is fixed. Thus, we will note a larger number of occupied subbands for cases where $\alpha<1$.
In Fig. (\ref{Energia_B}), we show the behavior of the energy states with respect to the magnetic field for some values of the $\alpha$ parameter.

\section{Magnetization}

The magnetization is given by
\begin{equation}
M = -\frac{\partial F}{\partial B}
= -\sum_{n,m}M_{n,m}f_{F}(E_{n,m}),  \label{Mag}
\end{equation}
where $F$ is free energy, $M_{n,m}\equiv-\partial E_{n,m}/\partial B$ defines the
magnetic moment, and $f_{F}(E_{n,m})$ is  the Fermi distribution function.
By using Eq. (\ref{Enm_ring}), the magnetic moment is written explicitly
as
\begin{equation}
	M_{n,m}=-\frac{\hbar e}{\mu \alpha^{2}}
	\left[
	\left(n+\frac{1}{2}+\frac{L}{2}\right)
	\frac{\omega_{c}}{\omega}-\frac{m-l}{2}
	\right].
	\label{Mnm}
\end{equation}
\vspace{0.2cm}
\begin{figure}[!ht!]
	\centering
	\includegraphics[width=\columnwidth]{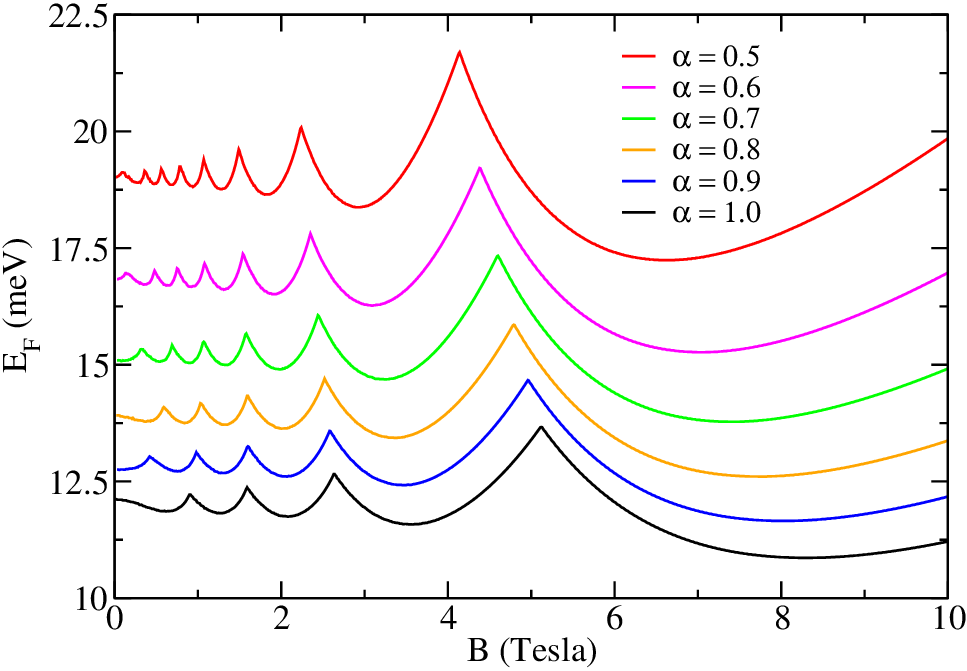}
	\caption{The zero temperature Fermi energy as a function of magnetic
	field.}
	\label{Fermi}
\end{figure}

In an isolated $2D$ ring, the number of electrons is constant, which induces a strong dependency of the Fermi energy with respect to the magnetic field. In the Fig. \ref{Fermi}, we display the behavior of the Fermi energy as a function of magnetic field strength. The oscillations result from the depopulation of a subband $n$. As already mentioned above, the $\alpha$ parameter increase the energy of the states, so an increase in the Fermi energy occurs. In addition, there is an evident increase in the number of occupied subbands.

\vspace{0.2cm}
\begin{figure}[!ht!]
	\centering
	\includegraphics[width=\columnwidth]{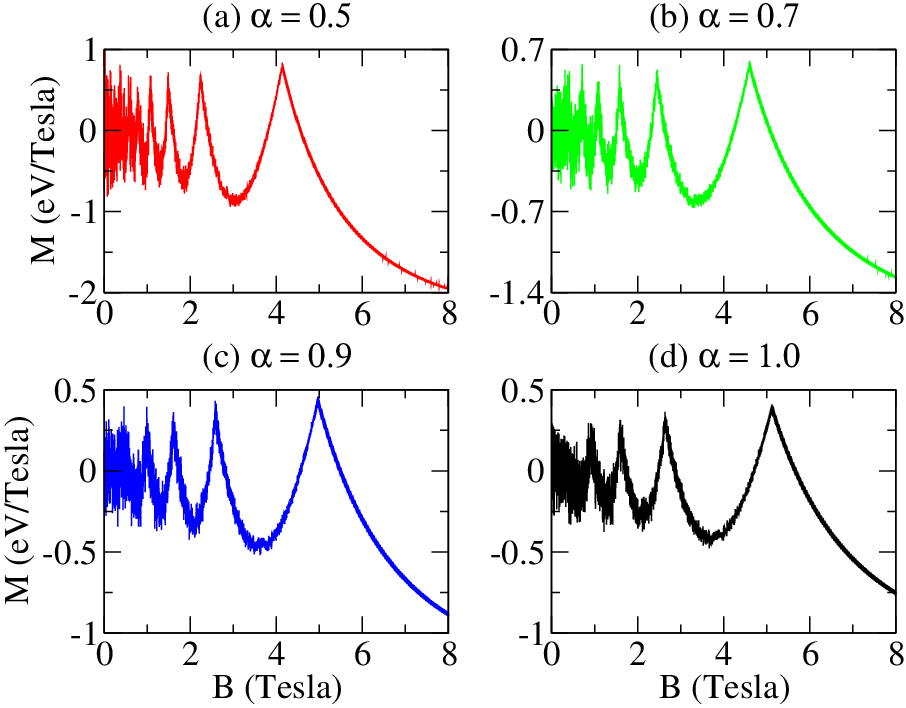}
	\caption{The magnetization of the quantum ring as a function of magnetic field
		strength.}
	\label{Magnetization}
\end{figure}

\begin{figure}[!ht!]
	\centering
	\includegraphics[width=\columnwidth]{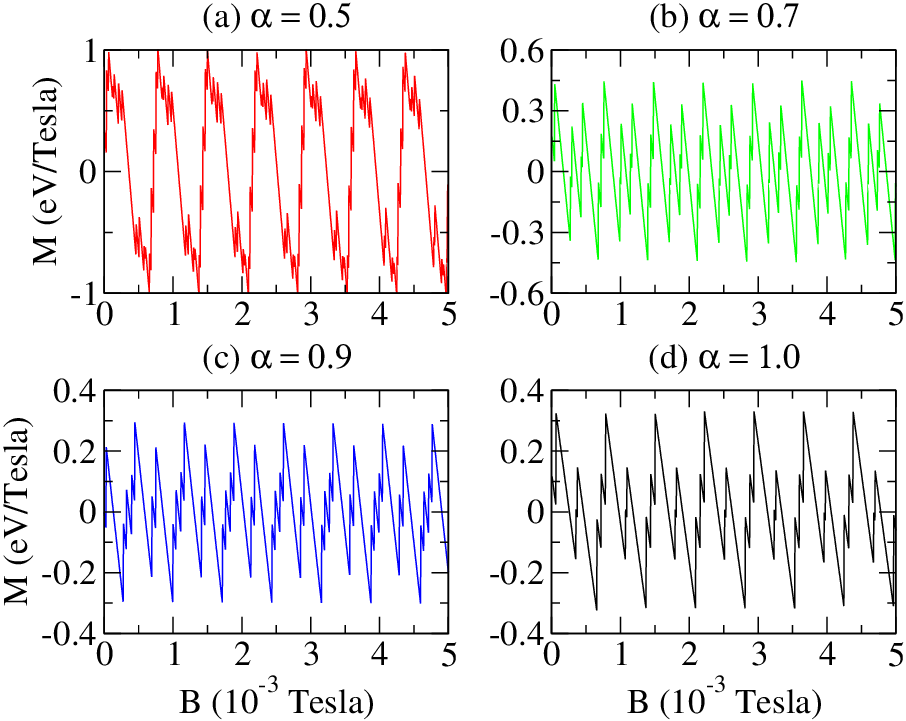}
	\caption{The magnetization of the quantum ring in the weak magnetic fields regime.}
	\label{Magnetization-weak}
\end{figure}

\begin{figure}[!ht!]
	\centering
	\includegraphics[width=\columnwidth]{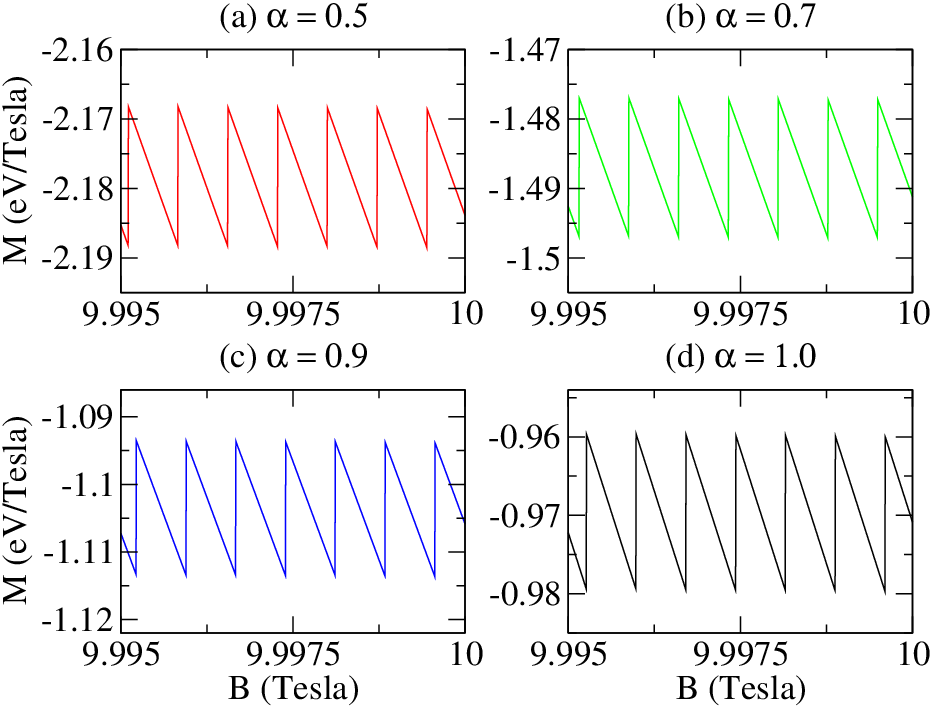}
	\caption{The magnetization of the quantum ring in the strong magnetic fields regime.}
	\label{Magnetization-strong}
\end{figure}

In Fig. \ref{Magnetization}, we show the behavior of the magnetization as a function of magnetic field. It is possible to see types AB and dHvA oscillations, with or without the presence of curvature. The AB type oscillations result from the states crossing, whereas the dHvA type oscillations result from the depopulation of subbands. In the weak magnetic fields regime (Fig. \ref{Magnetization-weak}), AB type oscillations are superimposed on dHvA type oscillations. For the $\alpha$ parameters considered, the periodic oscillations are maintain. In addition, we observe an increase in the amplitude of oscillations. By increasing the magnetic field, the dHvA type oscillations are more evident, whereas we see a decreasing of AB type oscillations amplitudes.
Also, we observe both a larger number of dHvA type oscillations, because of the increase of occupied subbands, and increase dHvA oscillations amplitudes when $\alpha<1$.
Even when there is only one occupied subband, we yet can see AB type oscillations (Fig. \ref{Magnetization-strong}).

We now show the shape of the oscillations when the temperature is non-zero in the weak magnetic field interval for the particular case when $\alpha=0.7$. As we can see in the Fig. {\ref{Magnetization-temperatura}, non-zero temperatures decrease the oscillations amplitude, and make the magnetization behavior a smooth function of the magnetic field.

\vspace{1.0cm}
\begin{figure}[!ht!]
	\centering
	\includegraphics[width=\columnwidth]{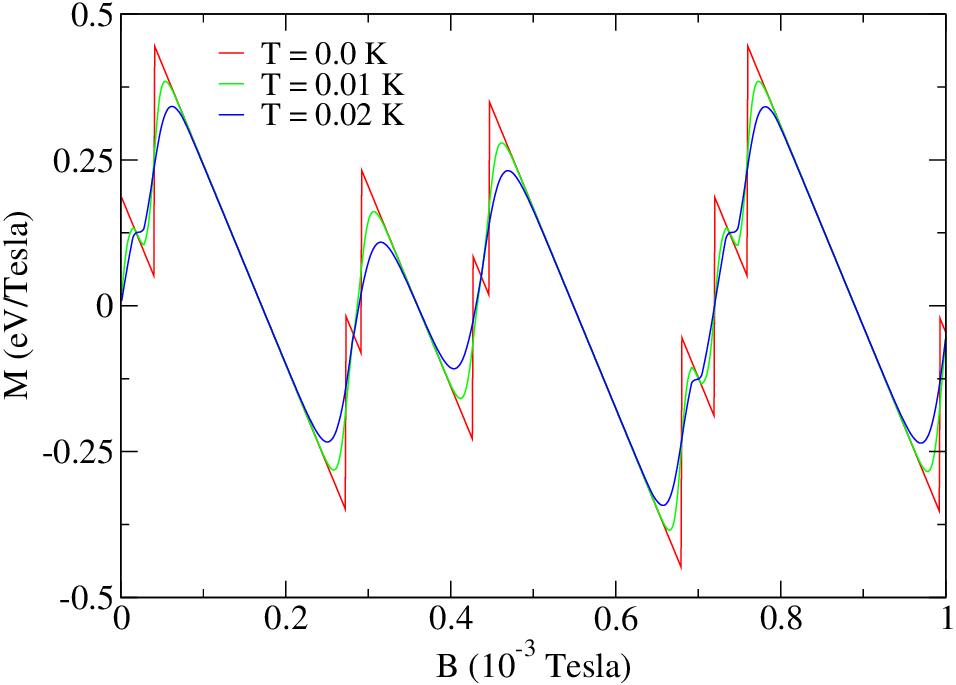}
	\caption{The effect of temperature on the oscillations observed in the magnetization for the case where $ \alpha = 0.7 $. We consider the interval of weak magnetic fields.}
	\label{Magnetization-temperatura}
\end{figure}

\section{Persistent current}

By considering the variation of the magnetic flux confined to the hole
of the ring, the persistent current is calculated using the
Byers-Yang relation \cite{PRL.7.46.1961}
\begin{equation}
	I_{n,m}=-\frac{\partial E_{n,m}}{\partial \Phi}=
	- \frac{1}{\phi_{0}} \frac{\partial E_{n,m}}{\partial l},
\end{equation}
where $E_{n,m}$ is then given by Eq. (\ref{Enm_ring}) with $r_0=0$.
The total current is given by
\begin{equation}
	I=\sum_{n,m}I_{n,m}f_{F}(E_{n,m}),  \label{Cor}
\end{equation}
where $I_{n,m}$ is explicitly given by
\begin{equation}
	I_{n,m}=\frac{e\omega}{4\pi \alpha ^{2}}\left( \frac{m-l}{L}-\frac{\omega_{c}}{\omega}\right),  \label{Inm}
\end{equation}
which is the persistent current carried by a given state $\chi_{n,m}$ of
the ring.

As noted by Tan and Inkson \cite{PRB.1999.60.5626}, the proportionality between current and the magnetic moment observed in a one-dimensional ring is broken in a two-dimensional ring as result of the penetration of the magnetic field into the conducting region of the ring. In the presence of curvature, this relation between the current and the magnetic moment is given by
\begin{equation}
	M_{n,m}\left( B\right) =\pi r_{n,m}^{2}I_{n,m}-\frac{e\hbar}{\mu \alpha^{2}
	}\left(n+\frac{1}{2}\right) \frac{\omega_{c}}{\omega}.
	\label{Mnm_Inm}
\end{equation}
The first term is the current in a one-dimensional ring with radius $r_{n,m}$ given by Eq. (\ref{raio_nm}).
The second term results from the penetration of the magnetic field into
the $2$D structure, being it a diamagnetic term.
From this result it is possible to show that if
$\omega_{c}\ll \omega_{0}$ the persistent current and the magnetization
present a similar behavior.
\vspace{0.2cm}
\begin{figure}[!ht!]
	\centering
	\includegraphics[width=\columnwidth]{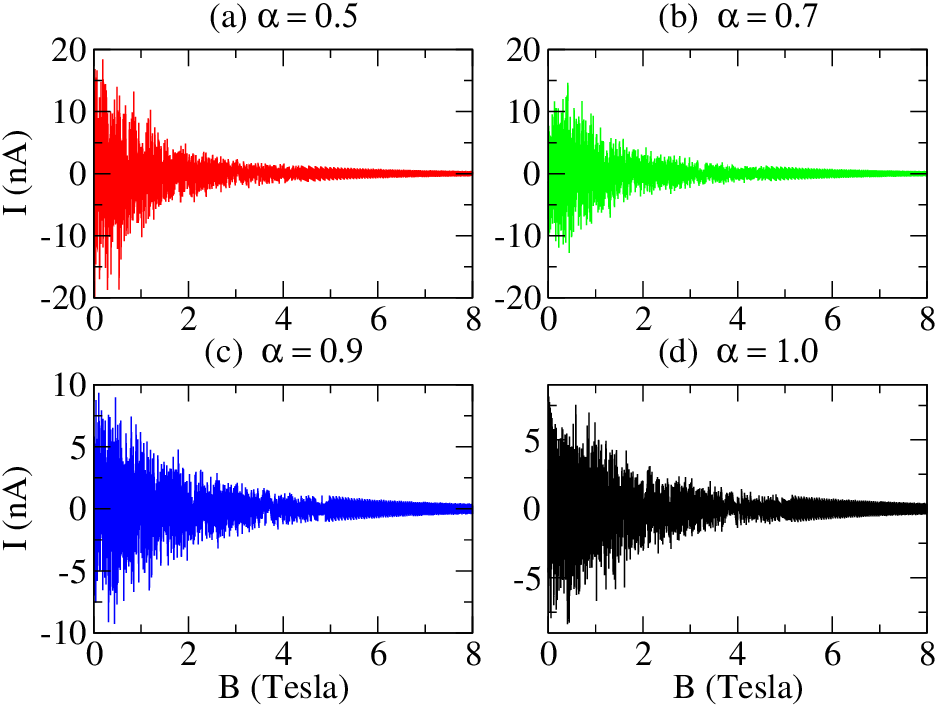}
	\caption{The persistent currents of the quantum ring as a function of the magnetic field strength.}
	\label{Corrente}
\end{figure}

\begin{figure}[!ht!]
	\centering
	\includegraphics[width=\columnwidth]{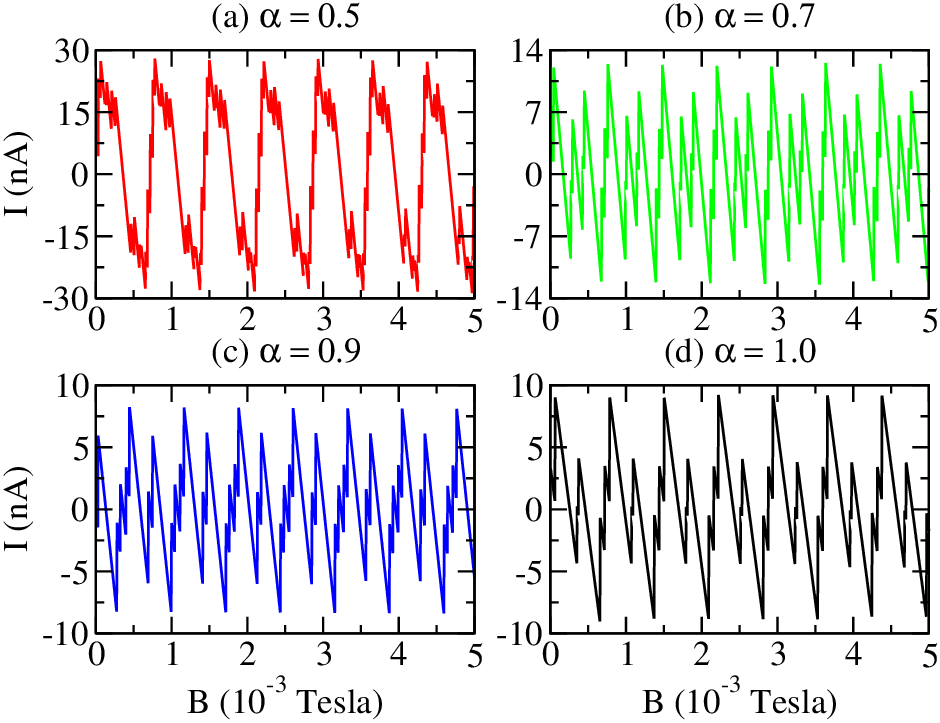}
	\caption{The persistent currents in the weak magnetic fields regime.}
	\label{Corrente-weak}
\end{figure}

\begin{figure}[!ht!]
	\centering
	\includegraphics[width=\columnwidth]{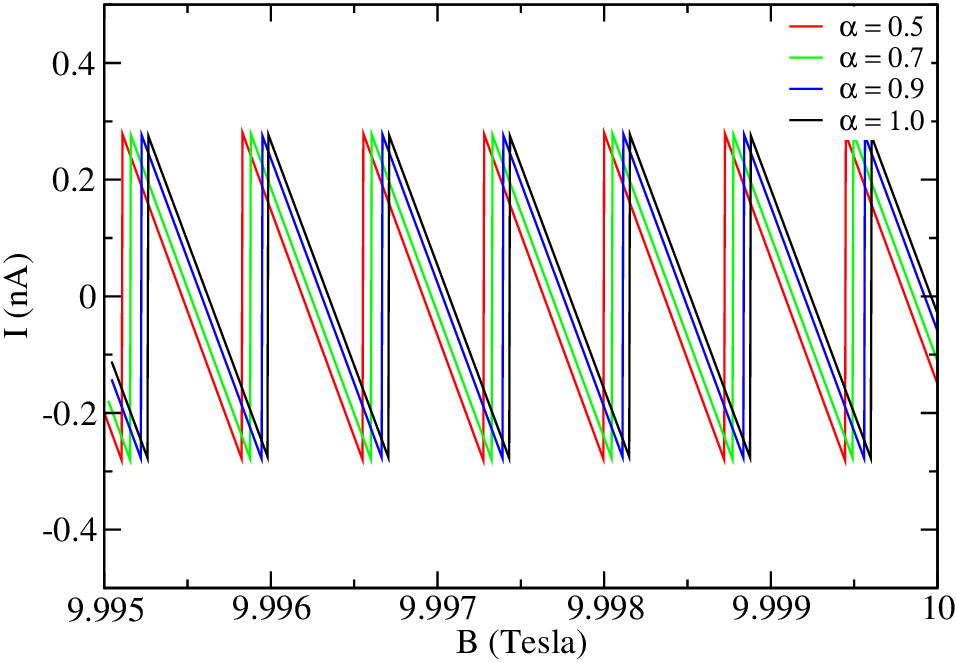}
	\caption{The persistent currents in the strong magnetic fields regime.}
	\label{Corrente-strong}
\end{figure}

In the Fig. (\ref{Corrente}), we show the behavior of the persistent current in a $2$D ring as a function of the magnetic field for some values of the $\alpha$ parameter. We can see AB type oscillations within the whole magnetic field range, regardless of the $\alpha$ parameter value.  These oscillations amplitude are strongly suppressed by increasing magnetic field strength, however, they increase with $\alpha$ parameter. In the weak magnetic field regime (Fig. (\ref{Corrente-weak})), these oscillations are almost periodic. As noted above, the profile of the oscillations of the persistent current in the weak magnetic field interval are similar to the magnetization. In Fig. \ref{Corrente-strong}, we show the behavior of the persistent current in the strong magnetic field regime. As we can see, the AB oscillations are almost periodic. We also note that curvature only shift the maximum of the oscillations, however the amplitude does not change.
\vspace{1.0cm}
\begin{figure}[!ht!]
	\centering
	\includegraphics[width=\columnwidth]{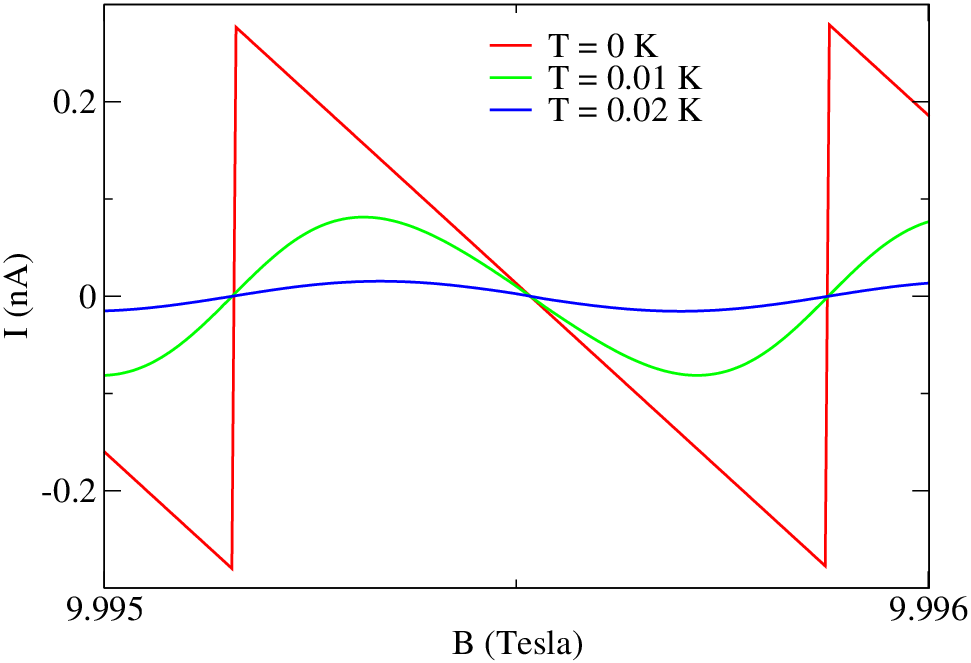}
	\caption{The effect of temperature on the oscillations observed in the persistent current for the case when $\alpha=0.7$. We consider a range of strong magnetic fields.}
	\label{Corrente-temperatura}
\end{figure}

In Fig. \ref{Corrente-temperatura}, we also show the shape of the oscillations on persistent current when the temperature is non-zero in the strong magnetic field regime for the case when $\alpha=0.7$. Likewise, as observed in the magnetization, non-zero temperatures decrease the oscillations amplitude, and make the persistent current behavior a smooth function of the magnetic field.

\begin{figure}[!ht!]
\includegraphics[width=\columnwidth]{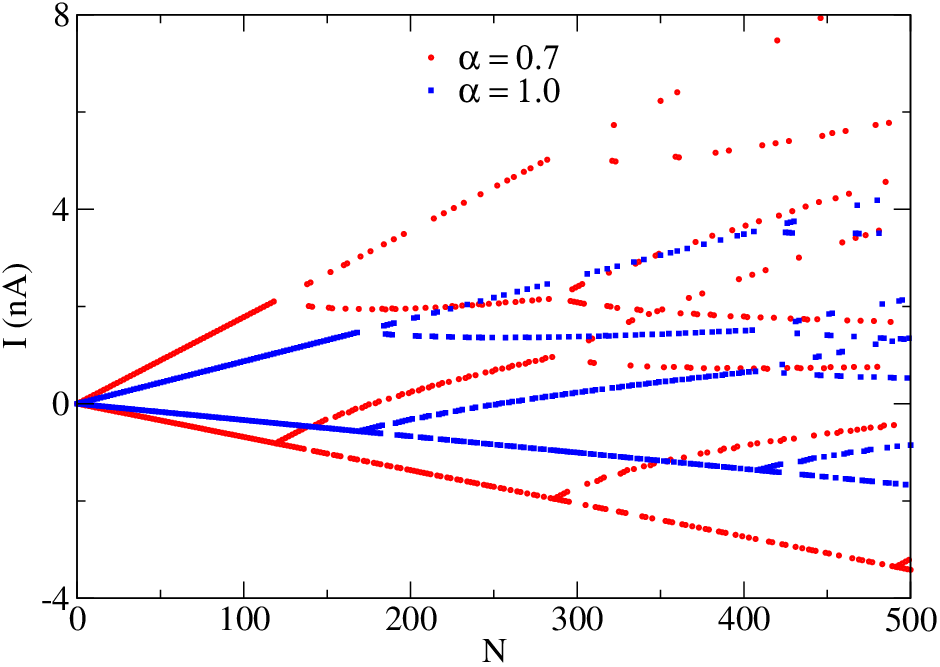}
\caption{Persistent current as a function of electron number for $\alpha=0.7$ e $\alpha=1.0$ parameters. We considered $B=1.0 \times 10^{-4}$  Teslas.}
\label{Corrente-dos}
\end{figure}

Finally, in Fig. \ref{Corrente-dos}, we show the persistent currents as a function of electron numbers, by considering a weak magnetic field regime, and the $\alpha=0.7$ and $\alpha=1.0$ parameter values.
We can note that the current exhibits oscillations in module and signal as a result of the occupation of the states that are to the right and to the left of the minimum of the subbands. This behavior is the result of the radial potential model used to describe the mesoscopic ring. The occupation of a new subband occurs whenever two new branches appear.

\section{Conclusions}
\label{Conc}

In this article, we studied the electronic properties, the
magnetization and the persistent current of a 2DEG in a
quantum ring in a conical metric and submitted to external
magnetic fields. We have obtained analytically the
wavefunctions and energy eigenvalues. It was shown that the curvature have strong influence on the spectrum of an electron in a quantum rings. The effects become more significant as the $\alpha$ parameter decreases.
In the numerical analysis of the magnetization, we observed AB and dHvA type oscillations, regardless of the $\alpha$ parameter. The curvature increased the amplitude of these oscillations, and also increased the number of dHvA type oscillations.
In the persistent currents, only AB type oscillations are observed, with an increase these amplitude for $\alpha<1$. It should be noted that the effect of the mean and Gaussian curvatures are negligible, because of the own geometry of the model. The same does not occur when geometry is a disk (quantum dot). As observed in the ref \cite{AdP.2019.531.1900254}, the mean curvature directly influences the behavior of the oscillations of magnetization and persistent currents.

\section*{Acknowledgments}

This work was partially supported by the Brazilian agencies CAPES, CNPq,
FAPEMA and FAPPR. FMA acknowledges CNPq Grants 313274/2017-7 and
434134/2018-0, and FAPPR Grant 09/2016. EOS acknowledges CNPq Grants
427214/2016-5 and 303774/2016-9, and FAPEMA Grants 01852/14 and 01202/16. This study was financed in part by the Coordena\c{c}\~{a}o de
Aperfei\c{c}oamento de Pessoal de N\'{\i}vel Superior - Brasil (CAPES) -
Finance Code 001.

\bibliographystyle{apsrev4-2}
%\bibliography{bibliography}
\input{curved-ring.bbl}

\end{document}